# ERROR STUDY OF CERN LINAC 4

M. Baylac*, JM de Conto, E. Froidefond, LPSC (CNRS/IN2P3-UJF-INPG), Grenoble, France,

E. Sargsyan, CERN, Geneva, Switzerland

*Abstract*

LINAC 4 is a normal conducting H- structure proposed to intensify the proton flux currently available for the CERN accelerator chain. This linac is designed to accelerate a 65 mA beam up to 160 MeV to be injected into the CERN Proton Synchroton Booster. The acceleration is performed up to 3 MeV by a Radio-Frequency Quadrupole resonating at 352 MHz followed by a series of two drift tube systems (conventional Alvarez and Cell Coupled Drift Tube Linac) boosting the beam up to 90 MeV at 352 MHz and finished by a Side Coupled Linac at 704 MHz. Beam dynamics was studied and optimized performing end-to-end simulations. Robustness of this design was verified by modelling machine errors. This paper presents the results of this error study.

## LINAC 4 LAYOUT

In the initial stage, LINAC 4 will be used as an injector to the PS Booster providing 40 mA average current of H- at 160 MeV with 0.08% duty cycle (d.c.). It is also conceived and designed as the normal conducting front-end of a 3.5 GeV superconducting proton linac with an average power 4-5 MW [1]. With such high beam power involved, beam quality must be controlled with extreme care to avoid activation and ensure hands-on operation. Although SPL d.c. will be 3-4%, the machine is designed for a 15% d.c.

LINAC 4 starts with a RF source, generating an H- beam at 95 keV. The first RF acceleration is done in a Radio-Frequency Quadrupole (the IPHI RFQ built at Saclay [2]). This 6 m long RFQ operating at 352 MHz brings the beam up to 3 MeV where it reaches the chopper line. A chopper is placed at 3 MeV to remove micro-bunches on the RF scale and rematch the beam to the following accelerating systems. The beam is then boosted to 40 MeV by a conventional Alvarez-type Drift Tube Linac (DTL) resonating at 352 MHz. The 13.4 m long DTL consists of 3 tanks and is fed by 5 klystrons. Beam focusing is performed in 82 cells with Permanent Magnet Quadrupoles. Further acceleration to 90 MeV is reached through a Cell-Coupled DTL at 352 MHz. The CCDTL consists of 72 cells powered by 8 klystrons. Electromagnetic Quadrupoles between the 24 tanks provide focusing. The final boost to 160 MeV is achieved via a Side Coupled Linac equipped with 20 Electromagnetic Quadrupoles, which resonates at 704 MHz. The SCL is made of 220 cells and is powered by 4 klystrons. Beam dynamics was studied and the design was finalised based on end-to-end simulations [3].

## ERROR STUDY

*Strategy*

The error study is performed on the 75 m long section of LINAC 4 including the DTL, the CCDTL and the SCL (see figure 1). The goal of this work is two-fold: define the manufacturing tolerances of the DTL, to be built in 2006, and examine the robustness of the LINAC 4 design as a whole. The RFQ tolerances have already been decided upon and the RFQ is now being built. The beam emittance used at the input of the DTL accounts for the RFQ output including errors. No correction scheme has been implemented.

This analysis is done in two stages. First, the sensitivity of the structure to one single error is determined in order to evaluate the individual contribution and fix an acceptable limit on each type of error. Then, all errors are combined simultaneously to verify the set of tolerances determined previously and estimate the overall degradation of the beam properties.

Simulations are performed with the Saclay code TraceWin [4]. Using its error module, we simulate alignment, focusing and RF errors, as follows:

- Quadrupole translations (transverse only, $\delta_x$, $\delta_y$) and rotations ($\phi_x$, $\phi_y$, $\phi_z$),
- Quadrupole gradient ($\Delta G/G$),
- Gap field ($\Delta E_{gap}/E_{gap}$),
- Klystron field and phase ($\Delta E_{klys}/E_{klys}$, $\phi_{klys}$).

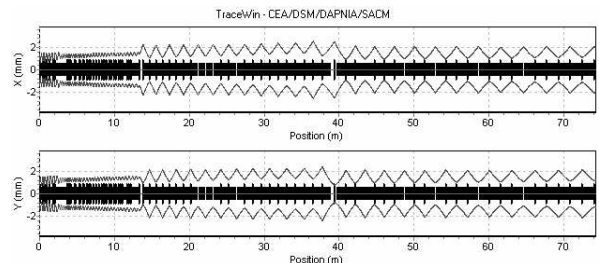

Figure 1: Beam envelope (5 RMS) through LINAC 4 DTL, CCDTL and SCL in x (top) and y (bottom).

Each error is applied on all linac cells. For each cell, the amplitude of the error is generated randomly and uniformly within a given range [-max, +max]. The relative emittance increase $\Delta\varepsilon$, in each run is expressed with respect to the nominal case, ie the case where beam is transported through the ideal linac without errors:

$$\Delta\varepsilon = \frac{\delta\varepsilon_{err} - \delta\varepsilon_{nom}}{\delta\varepsilon_{nom}}$$

*Corresponding author: baylac@lpsc.in2p3.fr

where $\delta\varepsilon_{err}$ and $\delta\varepsilon_{nom}$ are the emittance growth of the beam through the structure with and without errors. The natural transverse emittance growth in the nominal case $\delta\varepsilon_{nom}$ is ~9%. Each error simulation consists of 1000 runs. Beam loss and emittance growth are statistically averaged over the 1000 runs.

*Inputs*

At the entrance of the DTL, the beam has an energy of 3 MeV and its normalized RMS emittance is estimated to be $\varepsilon_x = \varepsilon_y = 0.28$ $\pi$.mm.mrad and $\varepsilon_z = 0.43$ $\pi$.mm.mrad. While the average current after chopping is 40 mA, the average current over the RF pulse is 65 mA and this is the intensity used in the error study simulations as it is the meaningful value for space charge effects. A Gaussian distribution with $5.10^4$ macro-particles per bunch is modelled in the first stage of this work. This number is increased to $10^6$ particles per bunch for the global simulations. Space charge interaction is calculated via the 3 dimensional PICNIC routine [5] with a 7x7 mesh, which is a good compromise between accuracy and calculation time.

*Individual sensitivities*

Figure 2 displays the statistical distribution of the calculated horizontal emittance increase with respect to the nominal case when all quadrupoles of the linac are shifted along the x direction by a random distance within [-0.1mm; +0.1mm]. For each of the nine types of errors defined above, we perform simulations while varying the maximum allowed amplitude of the error. This aims to determine the amplitude of each error minimizing beam degradation (no beam loss). As an example, figure 3 displays the average emittance increase with respect to the nominal case, if a random roll angle of varying maximum amplitude is applied to all the quadrupoles of the DTL. In this case, the generated emittance growth, similar along both transverse directions, rises quadratically with the roll angle. This behaviour is confirmed by independent theoretical calculations.

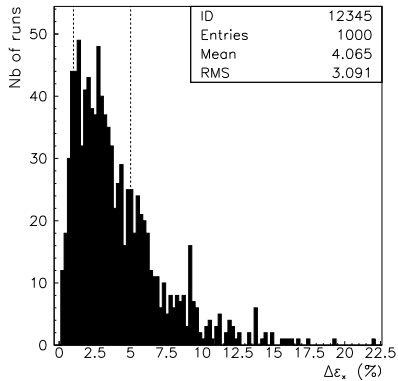

Figure 2: horizontal emittance increase when all linac quadrupoles are randomly shifted along x within ±0.1mm.

Discussions with RF and alignment experts along the study ensured that the tolerances obtained via simulations are achievable. Table 1 presents for all errors, the average and RMS of the relative emittance growth with respect to the nominal case, as well as the probability for $\Delta\varepsilon$ to be less than 1% or less than 5% in each simulation. Results are symmetric in x and y, such that for example, $\Delta\varepsilon_y \sim 4\%$ for $\delta_y \pm 0.1$ mm. No loss is detected within the quoted amplitudes.

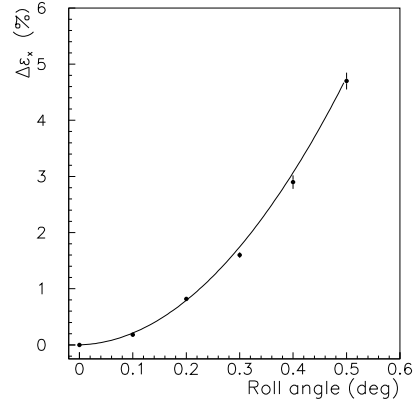

Figure 3: emittance growth when longitudinal rotations are applied to all DTL quadrupoles as a function of the maximum rotation amplitude (plotted for 45 mA). Superposed is a quadratic fit.

Table 1: sensitivities of the linac to errors

| Error type, amplitude | $\langle\Delta\varepsilon_x\rangle\pm$RMS probabilities | $\langle\Delta\varepsilon_y\rangle\pm$RMS probabilities | $\langle\Delta\varepsilon_z\rangle\pm$ RMS probabilities |
|---|---|---|---|
| $\delta_x$ ± 0.1 mm | 4.1 ± 3.1 < 1%: 10.1 < 5%: 70.2 | 1.1 ± 0.6 < 1%: 55.6 < 5%: 99.8 | 3.9 ± 2.7 < 1%: 7.8 < 5%: 73.9 |
| $\phi_x$ ± 0.5 deg | 0.0 ± 0.1 < 1%: 100 < 5%: 100 | 0.0 ± 0.1 < 1%: 100 < 5%: 100 | 0.0 ± 0.1 < 1%: 100 < 5%: 100 |
| $\phi_z$ ± 0.2 deg | 1.3 ± 1.3 < 1%: 53.1 < 5%: 98.4 | 1.7 ± 1.0 < 1%: 22.9 < 5%: 98.7 | 0.1 ± 0.1 < 1%: 100 < 5%: 100 |
| $\Delta G/G$ ± 0.5% | 0.5 ± 0.7 < 1%: 81.8 < 5%: 99.8 | 1.2 ± 1.0 < 1%: 49.5 < 5%: 99.0 | 0.1 ± 0.2 < 1%: 99.8 < 5%: 100 |
| $\Delta E_{gap}/E_{gap}$ ± 1% | 0.4 ± 0.7 < 1%: 79.8 < 5%: 99.9 | 0.6 ± 1.1 < 1%: 68.4 < 5%: 99.3 | 0.5 ± 1.3 < 1%: 67.4 < 5%: 99.7 |
| $\Delta E_{klys}/E_{klys}$ ± 1% | 1.9 ± 2.0 < 1%: 39.6 < 5%: 92.4 | 2.3 ± 2.8 < 1%: 43.3 < 5%: 84.4 | 3.5 ± 5.0 < 1%: 32.1 < 5%: 75.4 |
| $\phi_{klys}$ ± 1 deg | 1.4 ± 1.4 < 1%: 43.9 < 5%: 97.6 | 1.8 ± 2.0 < 1%: 41.2 < 5%: 91.9 | 3.0 ± 3.6 < 1%: 31.9 < 5%: 78.7 |

## RESULTS

*DTL tolerances*

After determining independently what seems to be an acceptable upper bound for each type of error, we verify their validity and estimate the total degradation of the beam properties using a global error simulation. This lengthy simulation (up to 400 CPU hours) with $10^6$ macro-particles per bunch combines all types of errors simultaneously.

At the beginning, we ran such global simulations limiting ourselves to the Drift Tube Linac, as this structure was to be manufactured first. The emittance increases in each direction by ~4% on average with respect to the nominal case when all errors applied. No particle loss is detected. We modified the input distribution and verified that the distribution of the emittance increase is simply shifted up or down by ~40% when modelling a Gaussian or a KV distribution. Thus we could fix the tolerances for the DTL to the values quoted in Table 1. These are comparable to the tolerances on other components of LINAC 4 (IPHI RFQ) or other accelerators (SNS). They were accepted by the manufacturer (ITEP-VNIIEF) and by CERN RF experts. The first DTL tank is presently under construction.

*Global error runs through LINAC 4*

Finally, global error simulations are run on the linac. Table 2 summarizes the results obtained when applying the nine errors within the DTL tolerances on the DTL, the CCDTL and the SCL. The sensitive parameters appear to be the quadrupole transverse alignment and longitudinal rotation. Moderate emittance increase is induced by klystron errors or errors on the quadrupole focusing gradient. Very little effect is due to errors on the accelerating field in the gaps or due to transverse quadrupole rotations. We see that the individual sensitivities roughly add up when combining different errors. This observation is useful as one can get a rough estimate of the overall beam degradation using sensitivity runs only, thus avoiding the lengthy global simulations. Under these conditions which account for a realistic linac structure, an average transverse emittance growth with respect to the nominal case is found to be on the order of 15% (see Table 2). In 18 out the 1000 runs, particles are lost along the linac. The estimated power lost is ~ 0.06 W/m along the 75 m of the DTL-CCDTL- SCL for a 15% d.c., which is well below the acceptable limit of 1 W/m.

Table 2: global error simulations of the linac

| $\langle\Delta\varepsilon_x\rangle\pm$RMS probabilities | $\langle\Delta\varepsilon_y\rangle\pm$RMS probabilities | $\langle\Delta\varepsilon_z\rangle\pm$ RMS probabilities | Lossy runs |
|---|---|---|---|
| 11.3 ± 5.1 < 5%: 6.1 < 15%: 79.9 < 30%: 99.2 | 13.3 ± 6.5 < 5%: 2.4 < 15%: 69.8 < 30%: 98.4 | 18.3 ± 11.9 < 5%: 4.1 < 15%: 46.9 < 30%: 90.0 | 18 out of 1000 |

## CONCLUSIONS

An error study was performed on the proposed CERN LINAC 4 (3 MeV to 160 MeV). It included an initial stage where the impact on the beam properties of quadrupole misalignment and gradient error, error on the accelerating field was determined. This led to the determination of the manufacturing and RF tolerances for the DTL, summarized as follows for the quadrupoles:
- Transverse displacements: $\delta_{x,y}=\pm 0.1$ mm
- Transverse rotations : $\phi_{x,y} = \pm 0.5$ deg
- Longitudinal rotations : $\phi_z = \pm 0.2$ deg
- Gradient: $\Delta G/G = = \pm 0.5$ %,

and for the accelerating field:
- Gap field: $\Delta E_{gap}/ E_{gap} = \pm 1\%$
- Klystron field $\Delta E_{klys}/ E_{klys} = \pm 1\%$
- Klystron phase $\phi_{klys}= \pm 1$ deg.

The most sensitive parameters were found to be the transverse alignment of the quadrupoles and their orientation around the beam axis.

Global simulations were then run with all errors combined simultaneously to verify tolerances and determine the overall beam degradation. The DTL tolerances were applied on the whole linac to estimate particle loss under realistic conditions. In our case, individual sensitivities to errors appear to be independent and roughly add up when combined. The beam quality was found to remain good: the emittance growth for all errors uncorrected is ~15% on average. We estimate the particle loss along the linac around 0.06 W/m for a 15% d.c., well below our acceptable limit.

This work thus confirms in addition to end-to-end simulations that the proposed design for LINAC 4 is robust and realistic. LINAC 4, although designed as a low d.c. machine to inject the CERN PS Booster, can also be used as an injector for a high power driver with a much higher d.c., as tolerances were determined assuming a 1 W/m loss limit at 15% d.c.

## ACKNOWLEDGEMENTS

We acknowledge the support of the European Community-Research Infrastructure Activity under the FP6 "Structuring the European Research Area" program (CARE, Contract No. RII3-CT-2003-506395).